%% file: paper.tex
\documentclass[conference]{IEEEtran}
\usepackage[utf8]{inputenc}
\usepackage{float}
\usepackage{listings}
\usepackage{dblfloatfix}
\usepackage{color}
\usepackage{multirow}
\usepackage{algorithm}
\usepackage{algorithmic}
\usepackage{tcolorbox}
\usepackage{multirow}
\usepackage{mathtools}
\usepackage{amsfonts}
\usepackage{xspace}
\usepackage{times}
\usepackage{array}
\usepackage[hyperindex,breaklinks]{hyperref}
\usepackage{url}
\usepackage{enumerate}
\usepackage{subcaption}
\usepackage[font=small,labelfont=bf]{caption}
\usepackage[normalem]{ulem}
\usepackage{pgfplots}
\usetikzlibrary{matrix}
\usepgfplotslibrary{groupplots}
\pgfplotsset{compat=newest}
\pgfplotsset{compat=newest,
legend style={font=\footnotesize},
label style={font=\footnotesize},
tick label style={font=\footnotesize},
title style={font=\footnotesize}}
\usepackage{color}
\usepackage{tikz}
\usetikzlibrary{shapes,decorations,arrows,calc,arrows.meta,fit,positioning}
\tikzset{
    -Latex,auto,node distance =1 cm and 1 cm,semithick,
    state/.style ={ellipse, draw, minimum width = 0.7 cm},
    point/.style = {circle, draw, inner sep=0.04cm,fill,node contents={}},
    bidirected/.style={Latex-Latex,dashed},
    el/.style = {inner sep=2pt, align=left, sloped}
}

\newcommand{\method}{\textsc{Airavata}}

\begin{document}
\pagestyle{plain}

\title{\method: Quantifying (Hyper) Parameter Leakage in Machine Learning}

\author{
    \IEEEauthorblockN{Vasisht Duddu\IEEEauthorrefmark{1}, D. Vijay Rao\IEEEauthorrefmark{2}}
    \IEEEauthorblockA{\IEEEauthorrefmark{1}Indraprastha Institute of Information Technology, Delhi, India}
    \IEEEauthorblockA{\IEEEauthorrefmark{2}Institute for Systems Studies and Analyses, Delhi, India}
    \IEEEauthorblockA{vduddu@tutamail.com, vijayrao@issa.drdo.in}
}

\maketitle

\begin{abstract}
Machine Learning models, extensively used for various multimedia applications, are offered to users as a blackbox service on the Cloud on a pay-per-query basis.
Such blackbox models are commercially valuable to adversaries, making them vulnerable to extraction attacks to reverse engineer the proprietary model thereby violating the model privacy and Intellectual Property.
Here, the adversary first extracts the model architecture or hyperparameters through side channel leakage, followed by stealing the functionality of the target model by training the reconstructed architecture on a synthetic dataset.
While the attacks proposed in literature are empirical, there is a need for a theoretical framework to measure the information leaked under such extraction attacks.
To this extent, in this work, we propose a novel \textit{probabilistic framework}, \method, to estimate the information leakage in such model extraction attacks.
This framework captures the fact that extracting the exact target model is difficult due to \textit{experimental uncertainty} while inferring model hyperparameters and \textit{stochastic nature of training} to steal the target model functionality.
Specifically, we use \textit{Bayesian Networks} to capture uncertainty in estimating the target model under various extraction attacks based on the \textit{subjective} notion of probability.
We validate the proposed framework under different adversary assumptions commonly adopted in literature to \textit{reason} about the attack efficacy.
Alternatively, we show that this can be viewed as the information gained (reduction in entropy) about the blackbox target model on performing different attacks.
This provides a practical tool to infer actionable details about extracting blackbox models and help identify the \textit{best} attack combination which \textit{maximises the knowledge} extracted (or information leaked) from the target model.
\end{abstract}


\input{intro}
\input{background}
\input{model_extraction}
\input{framework}
\input{evaluation}
\input{inftheory}
\input{related}
\input{conclusion}

{\footnotesize
\bibliographystyle{IEEEtranS}
\bibliography{paper.bib}
}

\end{document}

%% file: intro.tex
\section{Introduction}\label{introduction}

Machine Learning (ML) algorithms, specifically Deep Neural Networks, undergo an iterative design and development process to achieve state of the art performance on complex human level tasks such as speech recognition and object tracking and identification, using a variety of multimedia data.
Companies invest significant human resources and capital to design these Neural Networks making them an important Intellectual Property.
For instance, Amazon, Google, BigML, and Microsoft have adopted the business paradigm of ML as a Service (MLaaS).
The ML models are provided as a commercial service to customers on a pay-per-query basis which makes them commercially valuable to motivated adversaries.
An adversary, in such a setting, aims to train a substitute model with functionality and architecture close to the target model which are referred to as model extraction attacks.

In order to successfully mount a model extraction attack, the adversary's goal is to extract the knowledge stored in the model which includes both the \textit{hyperparameters (architecture)} and the \textit{weights or parameters (functionality)} of the target model.
For the specific case of extracting Deep Neural Networks, as considered in this work, the adversary is required to infer various model attributes (hyperparameters): number of layers, number of parameters in each layer and type of layers (maxpool, convolution, fully connected).
Once the adversary has an architecture similar to the target model, the next task is to steal the functionality of the model by training the substitute model through active learning on a synthetic dataset \cite{Chandrasekaran2018ExploringCB}\cite{orekondy2018knockoff}.
In a blackbox setting, such as MLaaS, the adversary can only query the target model through an API and get the corresponding output predictions.
Here, ML-based models can be trained to infer the target model hyperparameters based on the input-output pairs \cite{oh2018towards} or iteratively solving the equations for unknown parameter variables to extract the functionality \cite{orekondy2018knockoff}\cite{tramer2016stealing}.
However, assuming a stronger adversary with access to the hardware executing the model, side-channel leakage such as power consumption \cite{cryptoeprint:2018:477}, timing \cite{DBLP:journals/corr/abs-1812-11720} and cache side channels \cite{yan2018cache}\cite{hong2018security} can be exploited to infer details of confidential applications running on the hardware.
While the attacks proposed in literature are empirical, a theoretical framework to study the attacks and their efficacy is lacking.
In this work, we propose a theoretical framework to bridge the gap between the empirical attacks and a theoretical approach to evaluate the efficacy of various model extraction attacks.
Specifically, we address the following research question,
\begin{quote}
	\textit{How much knowledge about the target model can an adversary infer from model extraction attacks?}
\end{quote}
In other words,
\begin{quote}
  \textit{How much information does a Machine Learning model leak under different model extraction attacks?}
\end{quote}

\textbf{Motivation.} Deep Learning models are extensively used for various multimedia applications which are kept as a blackbox to protect against adversaries \cite{10.1145/3150226}.
The knowledge about the blackbox models enables an adversary to mount various other security and privacy attacks: generate adversarial examples \cite{Papernot:2017:PBA:3052973.3053009}, violate user's privacy \cite{shokri2017membership} and identify the inputs passed to the model \cite{wei2018know}\cite{Fredrikson:2015:MIA:2810103.2813677}.
This makes model extraction the first step to a wide range of attacks to violate the security of a multimedia system and privacy of user's data.
The problem of end-to-end model extraction attack: from inferring target model architecture to stealing its functionality, is viewed as the problem of maximising the information gained about the target blackbox model.
Here, each of the attack is a probe to identify a subset of hyperparameters about the target model which alleviates the adversary's knowledge about the blackbox.
The adversary benefits the most only by combining different attacks together available under the assumed threat model to infer maximum possible information about the target model.
Despite combining different attacks together, it is unlikely that the knowledge of the blackbox model is extracted with complete certainty.
In other words, it is very challenging to reconstruct and train a model with architecture and functionality \textit{exactly} same as the target model.
This arises due to the uncertainty from two major sources:
\begin{itemize}
\item Inferring and predicting the target model hyperparameters through different attacks are inherently stochastic and uncertain such as ML based attacks \cite{orekondy2018knockoff}\cite{NIPS2017_7219}.
\item Despite knowing the exact architecture, the reconstruction process to steal the model functionality (Knowledge Distillation and Active Learning) are stochastic, making it hard to steal the exact functionality \cite{Adam}\cite{44873}.
\end{itemize}
Both these factors add to the overall uncertainty in performing end-to-end model extraction attacks.
On performing a single attack, the adversary obtains incomplete knowledge about the target model.
Hence, the collective opinion of multiple attack vectors iteratively reduces the uncertainty from black box (incomplete knowledge) to white box (complete knowledge).
In order to quantify the knowledge extracted by the adversary or information leaked by the model under various model extraction attacks, we mathematically model these uncertainties using Bayesian Networks \cite{Heckerman2008}.

\textbf{\textit{Contributions.}} We propose a novel probabilistic framework, \method, to \textit{quantify the information leakage \textit{about} the model parameters and architecture} by capturing the uncertainty in performing various extraction attacks.
This uncertainty is captured using subjective probability distributions of different attacks and the inferred attributes of a Neural Network represented as random variables with cause-effect relationship within a Bayesian Network.
The total knowledge about the target model is computed as the conditional probability from the number of model attributes inferred \textit{given} that the adversary selects a combination of attacks.
Instead of using the frequentist approach of probability which requires large data, we use the \textit{subjective notion of probability} where the probability is defined as the \textit{belief} of occurrence of a particular event.
We experimentally validate our model by training a Bayesian Network on the data capturing the relation between different attacks and corresponding inferred hyperparameters.
The total knowledge extracted, representing the success of attacks, is computed using exact inference algorithm (Variable Elimination) on the joint probability distribution captured by the Bayesian Network.
This helps to analyze different possible combinations of attacks and reason about their effectiveness in extracting the target blackbox model.
Viewing these interactions of conflict between the attackers and defenders as a game, we identify the optimal attack combination of attacks to maximize and accordingly set up defences.

%% file: background.tex
\section{Background}\label{background}

\subsection{Machine Learning}

Given the space of data instances $X$ and space of corresponding truth labels $Y$, ML algorithms learn a classification function $f: X \longrightarrow Y$ that accurately map data samples in $X$ to its corresponding class in $Y$.
This is modelled as an optimisation problem where the parameters are computed by minimising the loss function $l(f(x), y)$ over each data instance $(x, y)$ as the difference in model's prediction $f(x)$ and the ground truth label $y$.
Instead of performing the optimisation on the entire data population $P(X, Y)$, we estimate the loss ($L_{D}$) over training dataset $D \subset X \times Y$ where each data point $(x,y)$ $\stackrel{i.i.d}{\sim}$ $D$.
However, ML models tend to overfit on the training data, i.e, the accuracy of the training data is much higher than the accuracy of evaluation (previously unseen) data~\cite{Bishop:2006:PRM:1162264}.
To ensure that the model does not overfit, a regularisation function ($J$) is added to the loss function $L_{D}$ which is balanced by the regularisation hyperparameter $\lambda$.
\begin{align}\label{eq:opt}
	\min\limits_f \, L_{D} + \lambda \, J
\end{align}

Neural Networks are a class of ML algorithms comprising of multiple computational units, called nodes, arranged in layers which are stacked sequentially.
Each layer performs matrix-vector multiplication between the updated parameter matrix and corresponding input activation from the previous layer.
This computation is followed by an activation function that restricts the output from growing too large.

\subsection{Probabilistic Graphical Models}

Complex systems are characterised by multiple inter-related attributes modelled as random variables to capture uncertainties in the system.
This can be modelled using the joint probability distribution over the set of random variables $\mathcal{X}$ but computing the entire distribution is computationally expensive, especially, for high dimensional networks with large number of variables.
Alternatively, Probabilistic Graphical Models encode the complex high dimensional joint probability distribution using a graphical structure \cite{Koller:2009:PGM:1795555}\cite{Heckerman2008}.
The nodes in the graphs represent the random variables characterising the complex system while the (lack of) edges between the nodes represent the conditional dependence or independence between the variables.

Bayesian Networks are a natural choice to model problems with uncertainty and causal relation between the variables \cite{Pearl:2009:CMR:1642718}.
Firstly, Bayesian Networks can learn from sparse incomplete datasets by probabilistically encoding dependencies between variables.
Secondly, they encode the causal relationship to help reason and infer about prior knowledge.
Further, Bayesian Networks provide declarative representation by encoding the system details while enabling algorithms to infer and reason about the knowledge captured in the models.

\textsc{Definition 1.} \textit{A Bayesian Network is a Directed Acyclic Graph $G=(V,E)$ with a random variable $x_i$ $\forall$ $i$ $\in$ $V$ characterised by a conditional probability distribution $p(x_i | p_{A}(x_i))$ per node specifying the probability of node $x_i$ conditioned on all the parent nodes $p_A$.}

Using the chain rule, the joint probability distribution $p(x_1, x_2, \dotsc, x_n)$ can be written as the product of $factors$,
\begin{equation}\label{joint}
p(x_1, \dotsc, x_n) = p(x_1) p(x_2 \mid x_1) \cdots p(x_n \mid x_{n-1}, \dotsc, x_1)
\end{equation}

Using Bayesian Networks for a problem has three components: (a) finding the graph structure to encode the causal relationship between variables from the data, (b) learning the conditional distributions between the variables and (c) identifying the probability of hypothesis variable.

\textbf{Structure Representation.} Structure learning algorithms \textit{search} the structure of the graphical model and identify the dependencies of different variables on each other using the data.
This is typically done by defining a score function, such as log-likelihood, to search among different DAGs for a structure which maximises the score using greedy or local search.
Alternatively, the constraint-based approach defines constraints on the edges of the graph and searches the optimal graph satisfying those constraints.
In this work, however, we build the graph structure based on our subjective knowledge about the domain due to the small number of variables.

\textbf{Parameter Learning.} Given the graph structure, we estimate the factors (conditional probabilities) for each of the nodes which make up the joint probability distribution.
In this work, we focus on Bayesian parameter estimation which explicitly models uncertainty over the node variables $x_i$ as well as the parameters of the Bayesian Network $\theta$ with a prior distribution $p(\theta)$ which encodes our subjective beliefs \cite{Krause:1993:RUK:563180}.
This deviates from the frequentist notion of probability which requires to enumerate all possibilities for a given hypothesis which is not possible for complex systems with an exponential number of cases and partial observability.
For each data point from the dataset $D$, the model updates its prior beliefs using the Bayes' rule,
\begin{equation}\label{bayes}
p(\theta \mid D) = \frac{p(D \mid \theta) \, p(\theta)}{p(D)}
\end{equation}
Here, the prior distribution assumed for the parameters $\theta$ is Dirichlet distribution which is iteratively updated based on new data samples.
This is specifically useful in our case of model extraction attacks where the data for training the Bayesian model is limited.

\textbf{Inference.} Given a Bayesian Network encoding dependencies between different random variables characterising the system, we estimate the probabilities of interest (inference).
The model encodes all probabilities to calculate all marginal, conditional and joint probabilities.
In this work, we focus on \textit{variable elimination algorithm} which is an exact inference algorithm.
Given the joint probability distribution across random variables $x_i$, we compute the marginal probability of $x_n$ by summing across all the other variables, i.e,
\begin{equation}\label{varel}
p(x_n) = \sum_{x_1} \cdots \sum_{x_{n-1}} p(x_1, \dotsc, x_n)
\end{equation}

%% file: model_extraction.tex
\section{Model Extraction Attacks}

The end-to-end model model extraction attack, viewed as a game between attacker and defender, is categorised into two problems: (a) inferring architecture and (b) stealing the functionality by training the substitute model ($f_{substitute}$) to have performance similar to target model ($f_{target}$).
In summary, the adversary aims to train a $f_{substitute}$ with both the architecture and functionality close to $f_{target}$.

\textbf{Attacker Knowledge.} For both architecture inference and stealing functionality: the $f_{target}$ is a blackbox, i.e, the architecture and parameter details of the model are not known to the adversary or any other user.
However, the strength of the adversary varies from having remote access to the target model to physical access to the hardware executing the model.
This results in leveraging different attacks with varying granularity and accuracy of the information extracted.

\subsection{Architecture Inference}

\textbf{Attacker Goal.} The first step of model extraction is to identify the architecture of $f_{substitute}$ as close as possible to $f_{target}$.
Though, it is possible to have different architectures with similar functionality, extracting $f_{substitute}$ to be close to $f_{target}$ improves the success of stealing the functionality as well as other security and privacy attacks to be mounted by the adversary.
The large number of hyperparameters of Deep Neural Networks makes the problem of identifying the exact $f_{target}$ challenging.
Hence, finding $f_{substitute}$ is modelled as a search problem, where the goal of the adversary is to leverage different attacks: side channels based and ML based, to reduce the overall search space \cite{DBLP:journals/corr/abs-1812-11720}.

\textbf{Attacker Action.} The attacker, to infer the hyperparameters, relies either on attacks exploiting cache, power, memory and timing side channels or using ML models to predict hyperparameters using input-output pairs of the target model.
Here, the attack varies from either performed remotely (timing and ML based attacks) or requiring hardware access (cache, power and memory side channel). These are considered within adversary models given in Section~\ref{advmodel}.

\textbf{Defender Action.} The goal of the defender is to protect against side channel leakage which is the primary approach for extracting architectural details.
Here, the defender can leverage various hardware optimisations to protect against timing side attack \cite{10.1145/3287624.3287694}, disallowing shared resources for cache attacks \cite{yan2018cache} and masking computation for power side channel leakage \cite{10.1007/978-3-642-38348-9}.

\subsection{Stealing Functionality}

\textbf{Adversary Goal.} Given a blackbox target Neural Network, the goal of the adversary is to train $f_{substitute}$ $\in$ $S$, where $S$ is the search space for all possible models with different hyperparameters, such that the functionality of $f_{substitute}$ approximates $f_{target}$ using minimum possible queries.
The test accuracy is used as a measure of performance and the objective function trains $f_{substitute}$ to minimize the difference between the predictions of $f_{target}$ and $f_{substitute}$ for inputs $(x,y)$ sampled from the data ($D$).
This ensures that the model resultant model makes predictions close to the target model predictions.

\textbf{Adversary Action.} The adversary has no knowledge about the underlying data and relies on creating a synthetic dataset to train $f_{substitute}$.
The assumptions about the adversary knowledge of the data and sampling mechanisms for creating the synthetic data from its underlying distribution, results in different stealing attacks with different performance \cite{DBLP:journals/corr/abs-1905-09165}\cite{orekondy2018knockoff}\cite{Chandrasekaran2018ExploringCB}\cite{Papernot:2017:PBA:3052973.3053009}.
In this work, all these variants are considered within a single node ``StealFunction" in the Bayesian Network which helps infer the parameters and hyperparameters in the objective function for $f_{substitute}$.

\textbf{Defender Action.} One proposed defence against stealing functionality of the model is based on identifying the difference in distribution between benign and adversarial queries \cite{DBLP:journals/corr/abs-1805-02628}.
Further, defences relying on reducing the granularity of the output predictions and adding an additional layer in the network to add noise have also been considered \cite{tramer2016stealing}\cite{orekondy2020prediction}\cite{Lee2018DefendingAM}.
Alternatively, stateful defences such as monitoring the information gained by a user with each query and raising an alarm when it exceeds a threshold has also been proposed \cite{10.1145/3274694.3274740}.

\subsection{Adversary Models}\label{advmodel}

Based on the above attacks for both architecture inference and functionality stealing, we consider three different adversary models enabling us with a tool to study and compare the effectiveness of different attacks specific to a particular setting.
Adversary models can be classified based on whether the attacker has physical access to the hardware running the target model or not.

\textbf{Adversary 1} (remote API interface) is weak and does not have physical access to the underlying target model hardware.
Instead, the adversary can query the target model through an API interface, where given an input image the attacker will receive the corresponding output prediction.
Here, the adversary can perform ML based architecture inference attack, timing side channel attack and functionality stealing attack since they do not require physical access to the hardware and can be performed remotely.

\textbf{Adversary 2} (physical access without API interface) has physical access to the underlying hardware.
For example, the adversary monitors the memory access pattern and measures the side channel information leakage including power side channel and cache accesses to infer architecture details.
Here, the adversary does not steal the functionality since there is no API for querying the model and this setting is used to evaluate the performance of only side channel leakage.

\textbf{Adversary 3} (physical access with API interface) has both API and physical access and hence can combine and evaluate the attacks executed by both Adversary 1 and Adversary 2.
In this case, the adversary can perform all the attacks to infer all possible information attributes about the target model.

\subsection{Sources of Uncertainty}

Uncertainty in model extraction occurs at two levels: (a) inferring model attributes by performing side channel based attacks and (b) reconstructing (training) the approximate architecture to have similar functionality as $f_{target}$ using stochastic learning algorithms.
While inferring model attributes, variability of experimental measurement while using ML models to infer target model attributes could result in \textit{experimental uncertainty} in both side channel based extraction attacks and ML based attacks.
Further, while training the attack models, there exists \textit{parameter uncertainty} where the model parameters are optimised and exact values are stochastic.
These inherent uncertainty in ML approaches \cite{NIPS2017_7219} can lead to imprecision in inferring the exact $f_{target}$.
Further, no single attack can extract the entire model with complete certainty and each attack infers only a subset of the overall model attributes.
Quantifying this uncertainty probabilistically determines the best possible performance of the $f_{substitute}$ that the adversary can achieve compared to $f_{target}$ based on the degree of knowledge that the attacker has inferred.

%% file: framework.tex
\section{\method\hspace{0.1cm} Framework}\label{framework}

\input{BN_fig}

In this section, we describe the proposed Bayesian Framework which represents various attacks and the inferred model attributes as random variables with a cause-effect relationship.
In other words, if an adversary chooses an attack, there exists a link between the attack variable to the corresponding attributes inferred in the Bayesian Network.
The structure of the Bayesian Network is designed based on the attacks and the corresponding attributes inferred proposed in literature.
This makes the framework practical and applicable to analyse the efficacy of real attacks in the literature.
Figure~\ref{fig:bayesnet} shows the proposed Bayesian Network with the attack nodes at the top layer followed by the inferred attributes and finally the knowledge of $f_{target}$ extracted by the adversary.

The model knowledge (last layer) is the hypothesis variable whose values are of interest to our problem.
The attack nodes (top layer) are the information variables for which the values are observed and influence the probability distribution of the hypothesis variable.
The information variables are linked to the hypothesis variable through intermediary variables (middle layer) which represent the inferred attributes.

\subsection{Attack Variables}

The modelling within Bayesian Framework categorises the attacks into different random variables (nodes in Bayesian Network) based on similarity of attack requirements (adversary models) and inferred attributes.
We consider that each attack node (information variable) represents the state of the art attack and are linked to the corresponding attributes inferred as described in the literature.

\textbf{Stealing Functionality (StealFunction).} Given large number of input-output pairs ($x,f(x)$), solving overdetermined system of equations for the unknown variables in terms of the known variables estimates the regularization hyperparameter from the objective function \cite{wang2018stealing}\cite{tramer2016stealing}.
Further, all the attacks falling within functionality stealing using active learning or retraining the model on synthetic data is considered within this variable \cite{orekondy2018knockoff}.
The node ``StealFunction" captures these attacks and enables to infer the hyperparameters used in the learning objective as well as estimate the values of the model parameters.

\textbf{ML against ML (MLvsML).} Machine Learning models can be trained to predict model attributes from the inputs-output predictions \cite{oh2018towards}.
Since, the attack uses ML models, they have uncertainty and error in predicting the model attributes correctly.
These attacks are abstracted within the ``MLvsML" node in the Bayesian Network and infer the number of layers, type of activation, number of parameters per layer and the type of layer.

\textbf{Timing Side Channel (TimingSC).} For a weak adversary with no knowledge about the target model, it is possible to infer the number of layers by computing the total execution time of the network \cite{DBLP:journals/corr/abs-1812-11720}.
The attack is based on the idea that all the nodes in a single layer are computed in parallel while all the layers are computed sequentially due to which the total execution time is strongly correlated to the number of layers.
Within the framework, this attack is captured in the node ``TimingSC" and infers only the number of layers in the Neural Network.

\textbf{Hardware Side Channel (HardwareSC).} An adversary with physical access to the hardware can monitor the memory access patterns during the model execution on the hardware (memory side channel) as well as exploit shared resources between processes to extract the process details (cache side channel) \cite{hong2018security}\cite{yan2018cache}.
Other hardware details like hardware performance counters, cache misses and data flow reveal significant internal model details \cite{DBLP:conf/ccs/Naghibijouybari18}\cite{Hua:2018:REC:3195970.3196105}\cite{DBLP:journals/corr/abs-1810-00602}\cite{unknown}.
All these attacks are abstracted as the ``HardwareSC" node and helps to infer the number of layers, type of activation, number of parameters per layer and the type of layer.
This is similar to ``MLvsML", however, the information inferred is more granular and accurate due to stronger adversary model.

\textbf{Power Side Channel (PowerSC).} During the execution of the neural network on hardware, a strong adversary with physical access to the target hardware, can monitor the power consumed to extract information about the application.
Given the power consumption traces, the attacker uses algorithms like differential power analysis, correlated power analysis and horizontal power analysis to infer the target black box model details \cite{cryptoeprint:2018:477}.
This is modelled as ``PowerSC" nodes within the framework and on successful execution, helps the adversary to infer the number of parameters in each layers, values of parameters, total number of layers and the type of activation function.

\subsection{Inferred Model Attributes}

Neural Networks have a large hyperparameter space where each hyperparameter can take different range of possible values.
The architecture details of the Neural Networks play a significant role in determining the performance.
Within \method, we model the major architecture attributes with significant influence on the performance of $f_{target}$.

\begin{itemize}
\item \textbf{Objective Function Hyperparameters (ObjHyperParam).} The objective function for training the Neural Network requires several hyperparameters such as the learning rate and momentum to control parameter updates, and weight decay to improve generalization.
The choice of loss functions along with the optimisation techniques determine the model performance.
\item \textbf{Number of Layers (Depth).} Typically, deeper the Neural Network the higher the performance due to which the ML community has focuses on scaling Neural Networks to large number of layers \cite{He2015DeepRL}\cite{Huang2016DenselyCC}.
\item \textbf{Parameters per Layer (Nodes).} The number of parameters per layer along with the model depth influences the complexity of the Neural Network which in turn affects the performance.
\item \textbf{Activation Function Type (Activation).} The type of activation function, $ReLU$, $Sigmoid$ or $Tanh$, maps the intermediate matrix-vector computation of each node to a range of output values.
\item \textbf{Layer Type.} The types of layers, convolutional, maxpool or fully connected, layer play an important role in determining the computation complexity and performance.
\end{itemize}

The causal relations between different nodes in the Bayesian Network enables one random variable to influence the probability distribution of other random variables.
In the proposed Bayesian Network, the V-Structure ($A \rightarrow B \leftarrow C$) indicates that the attribute random variables are independent of each other.
Hence, all attributes are independently inferred by an attack performed by the adversary.

\subsection{Extracted Knowledge}

The degree of knowledge extracted will be different for different attacks which the proposed model is required capture.
Formally, let the model attributes $M_i \in \mathcal{M}$ where $\mathcal{M}$=$\{M_1, \cdots, M_n\}$ and inferred attributes and the attacks $A_i \in \mathcal{A}$ where $\mathcal{A}$=$\{A_1, \cdots, A_m\}$.
The ultimate goal is to infer the hypothesis random variable, i.e, the degree of knowledge extracted $\mathcal{K}$.
The final knowledge is estimated as $P(\mathcal{K} | \mathcal{A})$ is the probability of the hypothesis random variable $\mathcal{K}$ given the evidence of the attack performed $\mathcal{A}$.
Training the Bayesian Network on data (described in Equation~\ref{bayes}) captures the joint probability distribution across all the random variables $P(\mathcal{K},\mathcal{M},\mathcal{A})$.
This distribution of $\mathcal{K}$ is then inferred using Variable Elimination algorithm as indicated in Equation~\ref{varel}.
In our case, the total number of inferred attributes $N(M)=6$ while the number of attack types considered $N(A)=5$.
The final knowledge extracted ($K$) is categorized into three classes based on the number of attributes influenced or linked ($N(M)$) on choosing different attack variables.

\[
    K=
\begin{cases}
    High,& \text{if } N(M) = 6\\
    Medium,  & \text{if } 3 \leq N(M) \leq 5 \\
    Low,  & \text{if } 0 \leq N(M) < 3\\
\end{cases}
\]

These are subjective thresholds chosen based on the the reduction of search space on inferring different attributes about the target model.
The attack variables in turn influence the probability distribution of the intermediate information variable (model attributes) influencing the final probability distribution of ``Model Knowledge".
The resultant probability of the hypothesis variable estimated after inference using Variable Elimination indicates the degree of knowledge about the model extracted by performing model extraction attacks.
By evaluating different combination of attacks one can probabilistically infer the model information leakage using on the knowledge acquired.

\textbf{Capturing Uncertainty.} Within the framework, each attack does not deterministically infer the final attribute, i.e, each attack is not assumes to have its intended effect with complete accuracy.
In other words, there is a probability associated with correctly inferring an attribute on choosing an attack.
This captures the uncertainty in model extraction attacks by associating a conditional probability over inferring an attack attribute \textit{given} a particular combination of attacks.
This probabilistic approach of estimating the attack success makes the framework a powerful tool for realistic settings with uncertainty.
The framework is scalable and can be extended to other novel attacks that are proposed in the literature with ease to study their attack efficacy.

%% file: BN_fig.tex
\usetikzlibrary{fit,backgrounds}
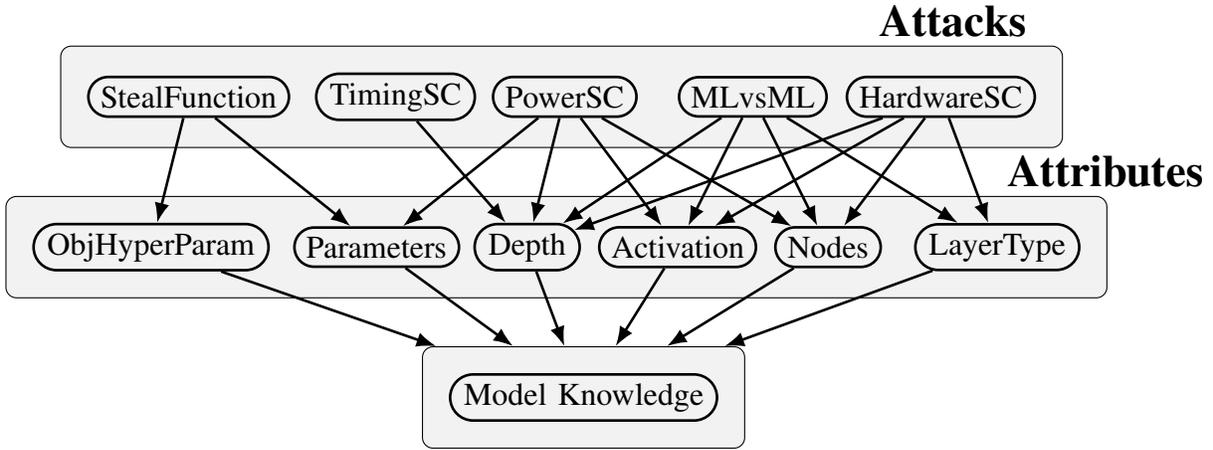
\begin{figure*}[ht!]
\centering
\begin{tikzpicture}[node distance=2cm, line width=1pt]

    \node[state, rounded rectangle] (eqsolve) at (-4.5,0) {\large StealFunction};
    \node[state, rounded rectangle] (timing) at (-1.75,0) {\large TimingSC};
    \node[state, rounded rectangle] (power) at (0.5,0) {\large PowerSC};
    \node[state, rounded rectangle] (blackbox) at (3,0) {\large MLvsML};
    \node[state, rounded rectangle] (map) at (5.5,0) {\large HardwareSC};

    \node[state, rounded rectangle] (hyperparameter) at (-5,-2) {\large ObjHyperParam};
    \node[state, rounded rectangle] (parameters) at (-2,-2) {\large Parameters};
    \node[state, rounded rectangle] (depth) at (0,-2) {\large Depth};
    \node[state, rounded rectangle] (activation) at (2,-2) {\large Activation};
    \node[state, rounded rectangle] (number) at (4,-2) {\large Nodes};
    \node[state, rounded rectangle] (type) at (6.25,-2) {\large LayerType};

    \node[state, rounded rectangle] (knowledge) at (0.75,-4) {\large Model Knowledge};

    \path (timing) edge (depth);
    \path (eqsolve) edge (hyperparameter);
    \path (eqsolve) edge (parameters);
    \path (power) edge (parameters);
    \path (power) edge (depth);
    \path (power) edge (activation);
    \path (power) edge (number);
    \path (blackbox) edge (depth);
    \path (blackbox) edge (type);
    \path (blackbox) edge (number);
    \path (blackbox) edge (activation);
    \path (map) edge (type);
    \path (map) edge (depth);
    \path (map) edge (number);
    \path (map) edge (activation);

    \begin{scope}[on background layer]
    \node[label={[xshift=5.2cm, yshift=0cm]\LARGE \textbf{Attacks}}, draw=black,rounded corners, solid,fit=(eqsolve) (power) (map) (blackbox) (timing), fill=gray!10, inner sep=0.35cm] (attacks) {};
    \node[label={[xshift=7.3cm, yshift=0cm]\LARGE \textbf{Attributes}},draw=black,rounded corners, solid,fit=(hyperparameter) (parameters) (activation) (depth) (number) (type), fill=gray!10, inner sep=0.35cm] (attributes) {};
    \node[draw=black,rounded corners, solid,fit=(knowledge) , fill=gray!10, inner sep=0.35cm] (final) {};
    \end{scope}

    \path (depth) edge (final);
    \path (type) edge (final);
    \path (parameters) edge (final);
    \path (hyperparameter) edge (final);
    \path (activation) edge (final);
    \path (number) edge (final);

\end{tikzpicture}
\caption{\method\hspace{0.1cm}Framework for Model Extraction Attacks in ML to estimate the knowledge extracted from the target model by the adversary.}
\label{fig:bayesnet}
\end{figure*}

%% file: evaluation.tex
\section{Evaluation}\label{evaluation}

In this section, we design and evaluate the Bayesian Network on data generated for model extraction attacks.
We implement the Bayesian Network using pgmpy\footnote{https://pgmpy.org/} library and design the structure of the Bayesian Network subjectively using expert knowledge as shown in Figure~\ref{fig:bayesnet}.
The code and the data used for experiments is available online for reproducing the results.

\subsection{Dataset}

The dataset captures the cause-effect relation between the different attacks, corresponding attributes and the overall knowledge inferred.
The data is compiled from real attacks and the corresponding inferred attributes based on the proposed literature to represent the real attack settings.
We consider the attack and corresponding attributes inferred to be binary variables, i.e, the attack can either be performed or not (``yes" or ``no") based on which the corresponding attributes are inferred.
This dataset considers five attack types: HardwareSC, PowerSC, MLvsML, TTimingSC and StealFunction attack.
The total attack combinations are hence, $2^5$ and the dataset has all the labels as discrete.
For each of the attack combinations, we have the corresponding inferred attributes as ``yes" while other attributes which are not inferred by the attack combination are labelled as ``no".
The final hypothesis variable is classified into ``High", ``Medium" and ``Low" based on the total number of inferred attributes on selecting a particular attack combination as described in Section~\ref{framework}.
While the dataset captures the attacks and their corresponding inferred variables as binary values, training of the Bayesian Network on the dataset updates the probability distributions to capture the probabilistic relation between different random variables.

\subsection{Inference: Adversary 1}

Given the Bayesian Network model capturing the joint probability distribution across the random variables, we query the model to reason about the effectiveness of different attacks based on the \textit{belief} of their extracted knowledge.
In the case of Adversary 1, we assume that the adversary is weak and has only remote API access to the target model.
In other words, the adversary can send queries (input images) to the target model and get the corresponding output predictions.
Here, the adversary can only perform attacks remotely and includes: TimingSC, MLvsML and StealFunction attack which can be mounted remotely according to their respective threat models.

\begin{table}[!htb]
\caption{Adversary 1: Belief in knowledge extracted in a remote black box setting with API.}
\begin{center}
\renewcommand\arraystretch{1.5}
\fontsize{6.7pt}{6.7pt}\selectfont
\begin{tabular}{|c|c|c|c|}
\hline
\multirow{2}{*}{\textbf{Attack Combination}}&\multicolumn{3}{c|}{\textbf{Knowledge Extracted (Leaked)}}\\
& Low & Medium & High\\ \hline
MLvsML &  0.0992 & 0.7983 & 0.1024\\
StealFunction &  0.7272 & 0.1586 &  0.1142\\
TimingSC &  0.7681 & 0.1178 &  0.1141\\
\hline
MLvsML + StealFunction &  0.0824 & 0.1822 &  0.7354\\
StealFunction + TimingSC &  0.1606 & 0.7262 &  0.1132\\
MLvsML + TimingSC &  0.0992 & 0.7983 &  0.1024\\
MLvsML + TimingSC + StealFunction &  0.0824 & 0.1822 &   0.7354\\
\hline
\end{tabular}
\end{center}
\label{adversary1}
\end{table}

The belief of knowledge extracted for remote black box setting corresponding to Adversary 1 is shown in Table~\ref{adversary1}.
In the remote setting, we can reason that since the number of attributes inferred by TimingSC and the functionality stealing attacks are less compared to MLvsML, the corresponding belief of extracting ``Low" knowledge is 0.7681 and 0.7272 respectively.
While for strong black box attacks like MLvsML, the knowledge extracted has been classified as ``Medium" with a belief score of 0.7983.
The StealFunction attack is typically performed after inferring some target model attributes to get an approximate architecture.
For the specific case of individual attack in Row 2 (Table~\ref{adversary1}), the consideration is for a random architecture chosen by the adversary.

However, the adversary best benefits from performing the attacks in combination rather than in isolation.
Specifically, we see that the adversary on combining all the three attacks has a belief score of 0.7354 for ``High" degree of knowledge extraction, i.e, likelihood of correctly inferring all the attributes in the network.
Interestingly, not performing TimingSC attack results in the same belief as TimingSC infers only the number of layers which is also inferred by MLvsML and hence, does not contribute to any further increase in the belief.

In summary, under Adversary 1, \textbf{the adversary's best attack combination is MLvsML with StealFunction to extract the maximum possible knowledge about the target model}.

\subsection{Inference: Adversary 2}

In case of Adversary 2, we assume a stronger adversary with physical access to the hardware executing the Neural Networks.
However, the adversary does not have an API access to query the model and hence, can analyse side channel based hyperparameter inference attacks.
Here, the adversary can perform hardware based side channel attacks such as cache side channels, memory access patterns and power side channels by monitoring the power consumed by the hardware during the execution of Neural Networks.

\begin{table}[!htb]
\caption{Adversary 2: Belief in knowledge extracted when adversary has physical access to the hardware.}
\begin{center}
\renewcommand\arraystretch{1.5}
\fontsize{6.7pt}{6.7pt}\selectfont
\begin{tabular}{|c|c|c|c|}
\hline
\multirow{2}{*}{\textbf{Attack Combination}}&\multicolumn{3}{c|}{\textbf{Knowledge Extracted (Leaked)}}\\
& Low & Medium & High\\ \hline
HardwareSC &  0.0992 & 0.7983 & 0.1024\\
PowerSC &  0.0894 & 0.8181 &  0.0925\\
\hline
HardwareSC + PowerSC &  0.0693 & 0.8142 &  0.1166\\
\hline
\end{tabular}
\end{center}
\label{adversary2}
\end{table}

Here, we see that the improvement in belief in using a combination of HardwareSC and PowerSC is not significant compared to performing the attacks independently (Table~\ref{adversary2}).
From this we can reason, that \textbf{both these attacks are equally strong in terms to extracting knowledge from the target model}.
However, on combining both the attacks we see that \textbf{the overall belief for ``High" knowledge increases from 0.1024 to 0.1166}.
However, the PowerSC has a higher belief for ``Medium" Knowledge extraction (0.8181 to 0.7983) compared to HardwareSC and MLvsML attack (remote adversary (Table~\ref{adversary1})).

\subsection{Inference: Adversary 3}

The third setting that we consider as part of our framework is where the adversary has physical access the hardware as well as the remote API to query the model.
This hypothetical setting allows to combine attacks from the above two settings to estimate the overall belief in extracting the target model knowledge.

\begin{table}[!htb]
\caption{Adversary 3: Belief in knowledge extracted when adversary has physical access to the hardware and remote API to query the target model.}
\begin{center}
\renewcommand\arraystretch{1.5}
\fontsize{6.7pt}{6.7pt}\selectfont
\begin{tabular}{|c|c|c|c|}
\hline
\multirow{2}{*}{\textbf{Attack Combination}}&\multicolumn{3}{c|}{\textbf{Knowledge Extracted (Leaked)}}\\
& Low & Medium & High\\ \hline
HardwareSC + StealFunction &   0.0824 & 0.1822 & 0.7354\\
PowerSC + StealFunction &  0.0893 & 0.7743 &  0.1364\\
HardwareSC + PowerSC + StealFunction &  0.0824 & 0.1822 &  0.7354\\
HardwareSC + MLvsML &  0.0992 & 0.7983 &  0.1024\\
PowerSC + MLvsML &  0.0693 & 0.8142 &  0.1166\\
\hline
All Attacks &  0.0824 & 0.1822 &  0.7354\\
\hline
\end{tabular}
\end{center}
\label{adversary3}
\end{table}

As shown in Table~\ref{adversary3}, combining different attacks together, results in a maximum belief of 0.7354 for extracting ``High" knowledge about target model.
However, the same level of knowledge can be inferred by choosing a careful combination of other attacks.
For instance, \textbf{HardwareSC + StealFunction and HardwareSC + PowerSC + StealFunction result in the same belief of extracting the overall knowledge instead of combining all the five attacks}.

%% file: inftheory.tex
\section{Discussion: Information Gain}

The subjective notion of probability used in Bayesian Networks can be viewed as reduction in entropy \cite{Das1999DSTORU}.
Various attacks performed on $f_{target}$ leak only a subset of attributes, reducing the overall entropy of the blackbox model.
This reduction in uncertainty due to model extraction attacks is the amount of information that the adversary acquires on performing various attacks.
Formally, consider $X$ and $Y$ as discrete random variables taking values ${x_1, \cdots ,x_n}$ with some probability $P(X=x_i)=P(x_i)$ and ${y_1, \cdots ,y_n}$ with some probability $P(Y=y_i)=P(y_i)$.
In order to specifically measure the uncertainty a particular attack reduces the blackbox model, we can compute the mutual information $I(X;Y)$ as,
\begin{equation}
I(X;Y)= \sum_{i} \sum_{j} P(x_i,y_j)log \frac{P(x_i,y_j)}{P(x_i)P(y_j)}
\end{equation}
The Mutual Information captures the uncertainty reduced in variable $X$ given random variable $Y$.
In other words, this is the information gained about the target model's knowledge $X$ given different attacks performed by the adversary $Y$.

\begin{table}[!htb]
\caption{Inferring attributes reduces the overall entropy of the target model.}
\centering
\renewcommand\arraystretch{1.5}
\fontsize{6.7pt}{6.7pt}\selectfont
\begin{tabular}{|c|c|}
\hline
\textbf{Attribute} & \textbf{Information Gain (bits)}\\
\hline
\multirow{1}{*}{Activation} & \multirow{1}{*}{0.372} \\
\multirow{1}{*}{Parameters/Layer} & \multirow{1}{*}{0.372} \\
\multirow{1}{*}{ObjHyperparam} & \multirow{1}{*}{0.556} \\
\multirow{1}{*}{Layer Type} & \multirow{1}{*}{0.346} \\
\multirow{1}{*}{Number of Layers} & \multirow{1}{*}{0.251} \\
\multirow{1}{*}{Parameters} & \multirow{1}{*}{0.227} \\
\hline
\end{tabular}
\label{tab:infgain}
\end{table}

The information gain for different parameters is given in Table~\ref{tab:infgain}.
This helps to understand how important different features are in the overall extraction of the knowledge.
Further, it provides the overall information gained by an adversary on inferring that particular attribute.
While this is specific to the attacks considered in the dataset, it helps to provide an additional tool to understand why some attacks perform better than others.
For instance, the attacks which are able to infer parameters with higher information gain will result in a higher leakage compared to other parameters.

%% file: related.tex
\section{Related Work}\label{related}

\textbf{Privacy Leakage in Machine Learning.} Machine Learning models leak information violating (a) data privacy through inference attacks and (b) model privacy through extraction attacks.
In data privacy, the goal of the adversary is to infer whether a given a data point is part of the training data or not by exploiting the difference in model's performance on training and testing data \cite{7958568}.
Alternatively, an adversary can infer attributes of training data which can further result in reconstruction attacks \cite{10.1145/3243734.3243834}.
In this work, the case of model extraction attacks is considered where the adversary aims to reverse-engineer the target black-box model.

\textbf{Applications of Bayesian Networks.} Bayesian Networks have been used extensively to model problems with inherent uncertainty.
This has a crucial application in cybersecurity threat detection where the uncertainty of the adversary's actions have to be taken into account \cite{inproceedings}\cite{6565234}\cite{Frigault:2008:MNS:1456362.1456368}.
A similar modelling can be done for network security attacks with various threats and exploits can be modelled as random variables in Bayesian Network \cite{5591406}.
Further, identifying data privacy risks by monitoring data access pattern and time duration can provide a tool based on Bayesian Networks to identify data privacy breaches \cite{An:2006:PID:1151454.1151493}.

\textbf{Quantifying Information Leakage.} Quantifying information leakage by measuring information flow from systems can help to identify and mitigate information leakage.
The state of the art approaches quantify information leakage about the \textit{input} based on the frequency of occurrence of input and outputs (frequentist approach) \cite{10.1007/978-3-319-11212-1_13}.
However, these mathematical measures compute the leakage measure from the conditional probabilities of outputs given the inputs but require large number of input-output data points \cite{Clark:2005:QIF:1094472.1094516}\cite{6266165}.
This requirement can be lifted by using machine learning models to quantify the leakage \cite{DBLP:journals/corr/abs-1902-01350}.
Orthogonal to work on quantifying information flow, this work focuses on modelling the black box systems itself and leakage corresponding to the model parameters and attributes using a Bayesian Networks (subjective approach of probability).

%% file: conclusion.tex
\section{Conclusions}\label{conclusion}

Model extraction attacks are a major threat to ML which violate the privacy and Intellectual Property of the proprietary algorithms deployed by companies.
In this work, we propose to unify such attacks under a single theoretical framework, \method, based on probabilistic graphical models which captures the uncertainty in performing model extraction attacks.
We model the attacks and inferred attributes as random variables with cause-effect relationship within the Bayesian Network and experimentally validate the framework under different adversary assumptions.
The inference of Bayesian Network enables to estimate the reduction of entropy (information gained) about the blackbox model.
The proposed framework enables to reason about the combination of attacks that maximise information leakage and provides a practical metric to compare the efficacy of various extraction attacks.


%% file: paper.bbl
\begin{thebibliography}{10}
\providecommand{\url}[1]{#1}
\csname url@samestyle\endcsname
\providecommand{\newblock}{\relax}
\providecommand{\bibinfo}[2]{#2}
\providecommand{\BIBentrySTDinterwordspacing}{\spaceskip=0pt\relax}
\providecommand{\BIBentryALTinterwordstretchfactor}{4}
\providecommand{\BIBentryALTinterwordspacing}{\spaceskip=\fontdimen2\font plus
\BIBentryALTinterwordstretchfactor\fontdimen3\font minus
  \fontdimen4\font\relax}
\providecommand{\BIBforeignlanguage}[2]{{%
\expandafter\ifx\csname l@#1\endcsname\relax
\typeout{** WARNING: IEEEtranS.bst: No hyphenation pattern has been}%
\typeout{** loaded for the language `#1'. Using the pattern for}%
\typeout{** the default language instead.}%
\else
\language=\csname l@#1\endcsname
\fi
#2}}
\providecommand{\BIBdecl}{\relax}
\BIBdecl

\bibitem{6266165}
M.~S. {Alvim}, K.~{Chatzikokolakis}, C.~{Palamidessi}, and G.~{Smith},
  ``Measuring information leakage using generalized gain functions,'' in
  \emph{2012 IEEE 25th Computer Security Foundations Symposium}, June 2012, pp.
  265--279.

\bibitem{An:2006:PID:1151454.1151493}
\BIBentryALTinterwordspacing
X.~An, D.~Jutla, and N.~Cercone, ``Privacy intrusion detection using dynamic
  bayesian networks,'' in \emph{Proceedings of the 8th International Conference
  on Electronic Commerce: The New e-Commerce: Innovations for Conquering
  Current Barriers, Obstacles and Limitations to Conducting Successful Business
  on the Internet}, ser. ICEC '06.\hskip 1em plus 0.5em minus 0.4em\relax New
  York, NY, USA: ACM, 2006, pp. 208--215. [Online]. Available:
  \url{http://doi.acm.org/10.1145/1151454.1151493}
\BIBentrySTDinterwordspacing

\bibitem{6565234}
E.~T. {Axelrad}, P.~J. {Sticha}, O.~{Brdiczka}, and J.~{Shen}, ``A bayesian
  network model for predicting insider threats,'' in \emph{2013 IEEE Security
  and Privacy Workshops}, May 2013, pp. 82--89.

\bibitem{cryptoeprint:2018:477}
L.~Batina, S.~Bhasin, D.~Jap, and S.~Picek, ``Csi neural network: Using
  side-channels to recover your artificial neural network information,''
  Cryptology ePrint Archive, Report 2018/477, 2018,
  \url{https://eprint.iacr.org/2018/477}.

\bibitem{Bishop:2006:PRM:1162264}
C.~M. Bishop, \emph{Pattern Recognition and Machine Learning (Information
  Science and Statistics)}.\hskip 1em plus 0.5em minus 0.4em\relax Berlin,
  Heidelberg: Springer-Verlag, 2006.

\bibitem{Chandrasekaran2018ExploringCB}
V.~Chandrasekaran, K.~Chaudhuri, I.~Giacomelli, S.~Jha, and S.~Yan, ``Exploring
  connections between active learning and model extraction.'' 2018.

\bibitem{DBLP:journals/corr/abs-1902-01350}
\BIBentryALTinterwordspacing
G.~Cherubin, K.~Chatzikokolakis, and C.~Palamidessi, ``{F-BLEAU:} fast
  black-box leakage estimation,'' \emph{CoRR}, vol. abs/1902.01350, 2019.
  [Online]. Available: \url{http://arxiv.org/abs/1902.01350}
\BIBentrySTDinterwordspacing

\bibitem{10.1007/978-3-319-11212-1_13}
T.~Chothia, Y.~Kawamoto, and C.~Novakovic, ``Leakwatch: Estimating information
  leakage from java programs,'' in \emph{Computer Security - ESORICS 2014},
  M.~Kuty{\l}owski and J.~Vaidya, Eds.\hskip 1em plus 0.5em minus 0.4em\relax
  Cham: Springer International Publishing, 2014, pp. 219--236.

\bibitem{Clark:2005:QIF:1094472.1094516}
\BIBentryALTinterwordspacing
D.~Clark, S.~Hunt, and P.~Malacaria, ``Quantitative information flow, relations
  and polymorphic types,'' \emph{J. Log. and Comput.}, vol.~15, no.~2, pp.
  181--199, Apr. 2005. [Online]. Available:
  \url{http://dx.doi.org/10.1093/logcom/exi009}
\BIBentrySTDinterwordspacing

\bibitem{Das1999DSTORU}
B.~Das, ``Dsto representing uncertainties using bayesian networks,'' 1999.

\bibitem{DBLP:journals/corr/abs-1812-11720}
\BIBentryALTinterwordspacing
V.~Duddu, D.~Samanta, D.~V. Rao, and V.~E. Balas, ``Stealing neural networks
  via timing side channels,'' \emph{CoRR}, vol. abs/1812.11720, 2018. [Online].
  Available: \url{http://arxiv.org/abs/1812.11720}
\BIBentrySTDinterwordspacing

\bibitem{10.1145/3287624.3287694}
\BIBentryALTinterwordspacing
A.~Fell, H.~T. Pham, and S.-K. Lam, ``Tad: Time side-channel attack defense of
  obfuscated source code,'' in \emph{Proceedings of the 24th Asia and South
  Pacific Design Automation Conference}, ser. ASPDAC ’19.\hskip 1em plus
  0.5em minus 0.4em\relax New York, NY, USA: Association for Computing
  Machinery, 2019, p. 58–63. [Online]. Available:
  \url{https://doi.org/10.1145/3287624.3287694}
\BIBentrySTDinterwordspacing

\bibitem{Fredrikson:2015:MIA:2810103.2813677}
\BIBentryALTinterwordspacing
M.~Fredrikson, S.~Jha, and T.~Ristenpart, ``Model inversion attacks that
  exploit confidence information and basic countermeasures,'' in
  \emph{Proceedings of the 22Nd ACM SIGSAC Conference on Computer and
  Communications Security}, ser. CCS '15.\hskip 1em plus 0.5em minus
  0.4em\relax New York, NY, USA: ACM, 2015, pp. 1322--1333. [Online].
  Available: \url{http://doi.acm.org/10.1145/2810103.2813677}
\BIBentrySTDinterwordspacing

\bibitem{Frigault:2008:MNS:1456362.1456368}
\BIBentryALTinterwordspacing
M.~Frigault, L.~Wang, A.~Singhal, and S.~Jajodia, ``Measuring network security
  using dynamic bayesian network,'' in \emph{Proceedings of the 4th ACM
  Workshop on Quality of Protection}, ser. QoP '08.\hskip 1em plus 0.5em minus
  0.4em\relax New York, NY, USA: ACM, 2008, pp. 23--30. [Online]. Available:
  \url{http://doi.acm.org/10.1145/1456362.1456368}
\BIBentrySTDinterwordspacing

\bibitem{10.1145/3243734.3243834}
\BIBentryALTinterwordspacing
K.~Ganju, Q.~Wang, W.~Yang, C.~A. Gunter, and N.~Borisov, ``Property inference
  attacks on fully connected neural networks using permutation invariant
  representations,'' in \emph{Proceedings of the 2018 ACM SIGSAC Conference on
  Computer and Communications Security}, ser. CCS ’18.\hskip 1em plus 0.5em
  minus 0.4em\relax New York, NY, USA: Association for Computing Machinery,
  2018, p. 619–633. [Online]. Available:
  \url{https://doi.org/10.1145/3243734.3243834}
\BIBentrySTDinterwordspacing

\bibitem{He2015DeepRL}
K.~He, X.~Zhang, S.~Ren, and J.~Sun, ``Deep residual learning for image
  recognition,'' \emph{2016 IEEE Conference on Computer Vision and Pattern
  Recognition (CVPR)}, pp. 770--778, 2015.

\bibitem{Heckerman2008}
\BIBentryALTinterwordspacing
D.~Heckerman, \emph{A Tutorial on Learning with Bayesian Networks}.\hskip 1em
  plus 0.5em minus 0.4em\relax Berlin, Heidelberg: Springer Berlin Heidelberg,
  2008, pp. 33--82. [Online]. Available:
  \url{https://doi.org/10.1007/978-3-540-85066-3_3}
\BIBentrySTDinterwordspacing

\bibitem{44873}
\BIBentryALTinterwordspacing
G.~Hinton, O.~Vinyals, and J.~Dean, ``Distilling the knowledge in a neural
  network,'' in \emph{NIPS Deep Learning and Representation Learning Workshop},
  2015. [Online]. Available: \url{http://arxiv.org/abs/1503.02531}
\BIBentrySTDinterwordspacing

\bibitem{hong2018security}
S.~Hong, M.~Davinroy, Y.~Kaya, S.~N. Locke, I.~Rackow, K.~Kulda,
  D.~Dachman-Soled, and T.~Dumitra{\c{s}}, ``Security analysis of deep neural
  networks operating in the presence of cache side-channel attacks,''
  \emph{arXiv preprint arXiv:1810.03487}, 2018.

\bibitem{unknown}
X.~Hu, L.~Liang, L.~Deng, S.~Li, X.~Xie, Y.~Ji, Y.~Ding, C.~Liu, T.~Sherwood,
  and Y.~Xie, ``Neural network model extraction attacks in edge devices by
  hearing architectural hints,'' 03 2019.

\bibitem{Hua:2018:REC:3195970.3196105}
\BIBentryALTinterwordspacing
W.~Hua, Z.~Zhang, and G.~E. Suh, ``Reverse engineering convolutional neural
  networks through side-channel information leaks,'' in \emph{Proceedings of
  the 55th Annual Design Automation Conference}, ser. DAC '18.\hskip 1em plus
  0.5em minus 0.4em\relax New York, NY, USA: ACM, 2018, pp. 4:1--4:6. [Online].
  Available: \url{http://doi.acm.org/10.1145/3195970.3196105}
\BIBentrySTDinterwordspacing

\bibitem{Huang2016DenselyCC}
G.~Huang, Z.~Liu, and K.~Q. Weinberger, ``Densely connected convolutional
  networks,'' \emph{2017 IEEE Conference on Computer Vision and Pattern
  Recognition (CVPR)}, pp. 2261--2269, 2016.

\bibitem{DBLP:journals/corr/abs-1805-02628}
\BIBentryALTinterwordspacing
M.~Juuti, S.~Szyller, A.~Dmitrenko, S.~Marchal, and N.~Asokan, ``{PRADA:}
  protecting against {DNN} model stealing attacks,'' \emph{CoRR}, vol.
  abs/1805.02628, 2018. [Online]. Available:
  \url{http://arxiv.org/abs/1805.02628}
\BIBentrySTDinterwordspacing

\bibitem{10.1145/3274694.3274740}
\BIBentryALTinterwordspacing
M.~Kesarwani, B.~Mukhoty, V.~Arya, and S.~Mehta, ``Model extraction warning in
  mlaas paradigm,'' in \emph{Proceedings of the 34th Annual Computer Security
  Applications Conference}, ser. ACSAC ’18.\hskip 1em plus 0.5em minus
  0.4em\relax New York, NY, USA: Association for Computing Machinery, 2018, p.
  371–380. [Online]. Available: \url{https://doi.org/10.1145/3274694.3274740}
\BIBentrySTDinterwordspacing

\bibitem{Adam}
\BIBentryALTinterwordspacing
D.~P. Kingma and J.~Ba, ``Adam: A method for stochastic optimization,'' 2014,
  cite arxiv:1412.6980Comment: Published as a conference paper at the 3rd
  International Conference for Learning Representations, San Diego, 2015.
  [Online]. Available: \url{http://arxiv.org/abs/1412.6980}
\BIBentrySTDinterwordspacing

\bibitem{Koller:2009:PGM:1795555}
D.~Koller and N.~Friedman, \emph{Probabilistic Graphical Models: Principles and
  Techniques - Adaptive Computation and Machine Learning}.\hskip 1em plus 0.5em
  minus 0.4em\relax The MIT Press, 2009.

\bibitem{5591406}
S.~{Kondakci}, ``Network security risk assessment using bayesian belief
  networks,'' in \emph{2010 IEEE Second International Conference on Social
  Computing}, Aug 2010, pp. 952--960.

\bibitem{Krause:1993:RUK:563180}
P.~Krause and D.~Clark, \emph{Representing Uncertain Knowledge: An Artificial
  Intelligence Approach}.\hskip 1em plus 0.5em minus 0.4em\relax Norwell, MA,
  USA: Kluwer Academic Publishers, 1993.

\bibitem{NIPS2017_7219}
B.~Lakshminarayanan, A.~Pritzel, and C.~Blundell, ``Simple and scalable
  predictive uncertainty estimation using deep ensembles,'' in \emph{Advances
  in Neural Information Processing Systems 30}, I.~Guyon, U.~V. Luxburg,
  S.~Bengio, H.~Wallach, R.~Fergus, S.~Vishwanathan, and R.~Garnett, Eds.\hskip
  1em plus 0.5em minus 0.4em\relax Curran Associates, Inc., 2017, pp.
  6402--6413.

\bibitem{Lee2018DefendingAM}
T.~Lee, B.~Edwards, I.~Molloy, and D.~Su, ``Defending against model stealing
  attacks using deceptive perturbations,'' \emph{ArXiv}, vol. abs/1806.00054,
  2018.

\bibitem{DBLP:conf/ccs/Naghibijouybari18}
H.~Naghibijouybari, A.~Neupane, Z.~Qian, and N.~B. Abu{-}Ghazaleh, ``Rendered
  insecure: {GPU} side channel attacks are practical,'' in \emph{{ACM}
  Conference on Computer and Communications Security}.\hskip 1em plus 0.5em
  minus 0.4em\relax {ACM}, 2018, pp. 2139--2153.

\bibitem{oh2018towards}
S.~J. Oh, M.~Augustin, M.~Fritz, and B.~Schiele, ``Towards reverse-engineering
  black-box neural networks,'' 2018.

\bibitem{orekondy2018knockoff}
T.~Orekondy, B.~Schiele, and M.~Fritz, ``Knockoff nets: Stealing functionality
  of black-box models,'' \emph{arXiv preprint arXiv:1812.02766}, 2018.

\bibitem{orekondy2020prediction}
\BIBentryALTinterwordspacing
------, ``Prediction poisoning: Towards defenses against {\{}dnn{\}} model
  stealing attacks,'' in \emph{International Conference on Learning
  Representations}, 2020. [Online]. Available:
  \url{https://openreview.net/forum?id=SyevYxHtDB}
\BIBentrySTDinterwordspacing

\bibitem{DBLP:journals/corr/abs-1905-09165}
\BIBentryALTinterwordspacing
S.~Pal, Y.~Gupta, A.~Shukla, A.~Kanade, S.~K. Shevade, and V.~Ganapathy, ``A
  framework for the extraction of deep neural networks by leveraging public
  data,'' \emph{CoRR}, vol. abs/1905.09165, 2019. [Online]. Available:
  \url{http://arxiv.org/abs/1905.09165}
\BIBentrySTDinterwordspacing

\bibitem{Papernot:2017:PBA:3052973.3053009}
\BIBentryALTinterwordspacing
N.~Papernot, P.~McDaniel, I.~Goodfellow, S.~Jha, Z.~B. Celik, and A.~Swami,
  ``Practical black-box attacks against machine learning,'' in
  \emph{Proceedings of the 2017 ACM on Asia Conference on Computer and
  Communications Security}, ser. ASIA CCS '17.\hskip 1em plus 0.5em minus
  0.4em\relax New York, NY, USA: ACM, 2017, pp. 506--519. [Online]. Available:
  \url{http://doi.acm.org/10.1145/3052973.3053009}
\BIBentrySTDinterwordspacing

\bibitem{Pearl:2009:CMR:1642718}
J.~Pearl, \emph{Causality: Models, Reasoning and Inference}, 2nd~ed.\hskip 1em
  plus 0.5em minus 0.4em\relax New York, NY, USA: Cambridge University Press,
  2009.

\bibitem{10.1145/3150226}
\BIBentryALTinterwordspacing
S.~Pouyanfar, Y.~Yang, S.-C. Chen, M.-L. Shyu, and S.~S. Iyengar, ``Multimedia
  big data analytics: A survey,'' \emph{ACM Comput. Surv.}, vol.~51, no.~1,
  Jan. 2018. [Online]. Available: \url{https://doi.org/10.1145/3150226}
\BIBentrySTDinterwordspacing

\bibitem{10.1007/978-3-642-38348-9}
E.~Prouff and M.~Rivain, ``Masking against side-channel attacks: A formal
  security proof,'' in \emph{Advances in Cryptology -- EUROCRYPT 2013},
  T.~Johansson and P.~Q. Nguyen, Eds.\hskip 1em plus 0.5em minus 0.4em\relax
  Berlin, Heidelberg: Springer Berlin Heidelberg, 2013, pp. 142--159.

\bibitem{7958568}
R.~{Shokri}, M.~{Stronati}, C.~{Song}, and V.~{Shmatikov}, ``Membership
  inference attacks against machine learning models,'' in \emph{2017 IEEE
  Symposium on Security and Privacy (SP)}, May 2017, pp. 3--18.

\bibitem{shokri2017membership}
R.~Shokri, M.~Stronati, C.~Song, and V.~Shmatikov, ``Membership inference
  attacks against machine learning models,'' in \emph{Security and Privacy
  (SP), 2017 IEEE Symposium on}, 2017.

\bibitem{DBLP:journals/corr/abs-1810-00602}
\BIBentryALTinterwordspacing
S.~Tople, K.~Grover, S.~Shinde, R.~Bhagwan, and R.~Ramjee, ``Privado: Practical
  and secure {DNN} inference,'' \emph{CoRR}, vol. abs/1810.00602, 2018.
  [Online]. Available: \url{http://arxiv.org/abs/1810.00602}
\BIBentrySTDinterwordspacing

\bibitem{tramer2016stealing}
F.~Tram{\`e}r, F.~Zhang, A.~Juels, M.~K. Reiter, and T.~Ristenpart, ``Stealing
  machine learning models via prediction apis,'' in \emph{USENIX Security},
  2016.

\bibitem{wang2018stealing}
B.~Wang and N.~Z. Gong, ``Stealing hyperparameters in machine learning,''
  \emph{arXiv preprint arXiv:1802.05351}, 2018.

\bibitem{wei2018know}
L.~Wei, Y.~Liu, B.~Luo, Y.~Li, and Q.~Xu, ``I know what you see: Power
  side-channel attack on convolutional neural network accelerators,''
  \emph{arXiv preprint arXiv:1803.05847}, 2018.

\bibitem{inproceedings}
P.~Xie, J.~Li, X.~Ou, P.~Liu, and R.~Levy, ``Using bayesian networks for cyber
  security analysis,'' 09 2010, pp. 211--220.

\bibitem{yan2018cache}
M.~Yan, C.~Fletcher, and J.~Torrellas, ``Cache telepathy: Leveraging shared
  resource attacks to learn dnn architectures,'' \emph{arXiv preprint
  arXiv:1808.04761}, 2018.

\end{thebibliography}
